\documentstyle[epsfig]{article}
\begin{document}
\centerline{ \Large{\bf{Lee-Yang zeros and the Ising model}}}
\centerline{ \Large{\bf{on the Sierpinski Gasket}}}
\vspace*{1cm}
\centerline{\large 
Raffaella Burioni\footnote{E-mail:raffaella.burioni@pr.infn.it},
Davide Cassi\footnote{E-mail:davide.cassi@pr.infn.it},
Luca Donetti\footnote{E-mail:donetti@prix7.fis.unipr.it}}
\centerline{\it 
Istituto Nazionale di Fisica della Materia, Unit\`a di Parma}
\centerline{\it 
Dipartimento di Fisica, Universit\`a di Parma}
\centerline{\it 
Parco Area delle Scienze 7a, 43100 Parma, Italy.}
\vspace*{1cm}

\begin{abstract}
We study the distribution of the complex temperature zeros for
the partition function of the Ising model on a Sierpinski gasket using an
exact recursive relation. Although the zeros arrange on a curve pinching the
real axis at $T=0$ in the thermodynamic limit, their density vanishes
asymptotically along the curve approaching the origin. This 
phenomenon explains the coincidence of the low temperature regime 
on the Sierpinski gasket and on the linear chain.
\end{abstract}

\section{Introduction}

The understanding of phase transitions on non-crystalline structures has
been recently improved by exact results connecting general geometrical features
of networks to the existence of spontaneous symmetry breaking 
\cite{panza1,panza2,mwg2,fss}.
The situation is  more complex when dealing with the critical behaviour of model systems.
For continuos symmetry models the singularity of the free energy, which determines the
critical behavior, appears to be related to the infrared spectrum of the
Laplacian operator on the network \cite{sfer,on}, while this is not the case for     
discrete symmetries (e.g. the Ising model). 
There, all known results suggest that the link between critical behaviour and geometry
should involve some other topological features \cite{panza1,panza2}.
An interesting result concerns the Sierpinski Gasket,
a typical and widely studied fractal, where the
Ising model is exactly solved. On this structure, although continuous symmetry models
exhibit a power law  behaviour for $T\to 0$, the Ising model has an exponential  low temperature behaviour which coincides with that found on the linear
chain \cite{panza1,panza2}.
To analyze the critical regime from an {\it ab initio} point of view, an interesting
picture is provided by the study of the singularities of thermodynamic
potentials. 
In 1952 Lee and Yang in two famous papers \cite{leeyang1,leeyang2} first
proposed their fundamental approach to phase transitions, consisting in studying
the zeros of the partition function of a statistical system,
considered as a function of a complex
parameter. The partition function on a finite volume is a
polynomial in complex activity or fugacity, so that the complete
knowledge of the zeros distribution is equivalent to the knowledge of
the partition function itself and all thermodynamic quantities can be
obtained from it. On a finite volume there are no real zeros, the coefficient
being all real and positive. However in the thermodynamic limit the zeros
can pinch the real axis (the region of physical interest) producing a 
singularity in the free energy (or grand-canonical potential)\cite{ruelle}.
The pinching points are phase transitions points on the
parameter axis and the zeros distribution in their neighbourhood can be
connected with the critical properties of the system \cite{ipz}.

Unfortunately, a complete knowledge of the zeros is very
difficult to obtain except for a few exactly solvable cases.
General theorems hold for the zeros distribution in the
complex magnetic field plane in a class of ferromagnetic
lattice systems, including the Ising model \cite{leeyang2,griff}.
On the other hand, very little
is known rigorously about the behaviour of the zeros of the partition function
in the complex temperature plane, the so called Fisher zeros \cite{fisher}.
In general, it is not clear if Fisher zeros arrange on smooth curves
even if this is the case in some exactly solvable models.
For the Ising model on regular two dimensional lattices \cite{fisher,shrock}, 
the Fisher zeros arrange on curves that cross the positive real axis at the 
transition point. In the one dimensional case only two zeros
(with infinite multiplicity) are found and these have a nonzero imaginary
part, so that there is no singular point for the free energy.

For statistical models defined on non periodic discrete structures,
Fisher zeros show some peculiar features, making the analysis
of their density and location extremely subtle.
In particular, on some hierarchical lattices (i. e. q-potts model 
on diamond hierarchical lattices \cite{derrida}) the zeros
have been show to form a fractal set (Julia sets). In this case, while
the general approach for identifying the singularity points and the
critical behaviour still holds,
the widely used arguments concerning scaling of singularities 
and zeros density with the volume must be handled carefully, 
as will be shown in the following.

In this paper we will study the Ising model on the Sierpinski gasket,
obtaining a recursive relation for the
partition function, from which the zeros of the $n$-th stage gasket can be
obtained from
those of the $(n-1)$th. The distribution we obtain is fractal and
pinches the real axis at $T=0$, so that a singular point with a power law
critical behaviour could be expected. However, since the zeros
density is found out to vanish exponentially in the neighbourhood of
$T=0$, these zeros don't produce any singularity
of the free energy: although the zeros pinch the real axis the `critical
behaviour' is the same as the one-dimensional case.

\section{Ising model on the Sierpinski gasket}
\begin{figure}[t]
  \begin{center}
  \begin{picture}(290,140)(-20,-20)
    \put(0,0){\begin{picture}(30,30) 
                      \multiput(0,0)(30,0){2}{\circle*{6}}
                      \put(15,30){\circle*{6}}
                      \put(0,0){\line(1,0){30}}
                      \put(0,0){\line(1,2){15}}
                      \put(15,30){\line(1,-2){15}}
                     \end{picture}}
    \put(10,-20){${\cal G}_0$}
    \put(50,0){\begin{picture}(60,60)
               \multiput(0,0)(30,0){2}{\begin{picture}(30,30)
                                          \multiput(0,0)(30,0){2}{\circle*{6}}
                                          \put(15,30){\circle*{6}}
                                          \put(0,0){\line(1,0){30}}
                                          \put(0,0){\line(1,2){15}}
                                          \put(15,30){\line(1,-2){15}}
                                       \end{picture} }
               \put(15,30){\begin{picture}(30,30) 
                              \multiput(0,0)(30,0){2}{\circle*{6}}
                              \put(15,30){\circle*{6}}
                              \put(0,0){\line(1,0){30}}
                              \put(0,0){\line(1,2){15}}
                              \put(15,30){\line(1,-2){15}}
                           \end{picture}}
               \end{picture}}
    \put(75,-20){${\cal G}_1$}
    \multiput(130,0)(60,0){2}{\begin{picture}(60,60)
                              \multiput(0,0)(30,0){2}{\begin{picture}(30,30) 
                      \multiput(0,0)(30,0){2}{\circle*{6}}
                      \put(15,30){\circle*{6}}
                      \put(0,0){\line(1,0){30}}
                      \put(0,0){\line(1,2){15}}
                      \put(15,30){\line(1,-2){15}}
                     \end{picture}}
                              \put(15,30){\begin{picture}(30,30) 
                      \multiput(0,0)(30,0){2}{\circle*{6}}
                      \put(15,30){\circle*{6}}
                      \put(0,0){\line(1,0){30}}
                      \put(0,0){\line(1,2){15}}
                      \put(15,30){\line(1,-2){15}}
                     \end{picture}}
                              \end{picture}}
    \put(160,60){\begin{picture}(60,60)
                   \multiput(0,0)(30,0){2}{\begin{picture}(30,30) 
                      \multiput(0,0)(30,0){2}{\circle*{6}}
                      \put(15,30){\circle*{6}}
                      \put(0,0){\line(1,0){30}}
                      \put(0,0){\line(1,2){15}}
                      \put(15,30){\line(1,-2){15}}
                     \end{picture}}
                   \put(15,30){\begin{picture}(30,30) 
                      \multiput(0,0)(30,0){2}{\circle*{6}}
                      \put(15,30){\circle*{6}}
                      \put(0,0){\line(1,0){30}}
                      \put(0,0){\line(1,2){15}}
                      \put(15,30){\line(1,-2){15}}
                     \end{picture}}
                  \end{picture}}
    \put(185,-20){${\cal G}_2$}
  \end{picture}
  \end{center}
  \caption{First iterations of gasket's construction}
  \label{gask}
\end{figure}
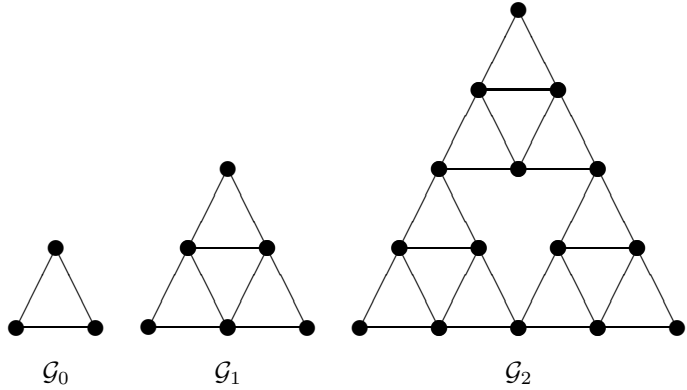

The Sierpinski gasket is a fractal graph which can be built recursively with
the following procedure: the initial stage (${\cal G}_0$) is a triangle 
(3 sites with 3 edges) and the $n$th stage (${\cal G}_n$) is obtained 
joining 3 ${\cal G}_{n-1}$
at their external corners, to form a bigger triangle (fig. \ref{gask}).
In this way ${\cal G}_n$ has $\frac{3}{2}(3^n-1)$ sites, $3^{n+1}$ edges and
its side contains $2^n$ edges.

The gasket is obtained as the limit for $n\rightarrow\infty$ of this procedure.

The Ising model on the gasket is defined associating the spin variable
$\sigma_i=\pm1$ to every site $i$ of the graph, and considering a 
nearest-neighbours interaction between points joined by an edge (link). 
The Hamiltonian is therefore:
\begin{equation}
E= -J\sum_{<i,j>}\sigma_i \sigma_j 
\end{equation}
where the sum runs over the couples of sites joined by a link and $J$ is a
positive constant 
(ferromagnetic coupling).

\section{Recursive relation for the partition function}

For ${\cal G}_0$ the partition function 
\begin{equation} 
Z=\sum_{\{\sigma_i \}} e^{- \beta E} 
\end{equation}
can be seen as a sum of the elements of the rank 3 tensor $M_0$
\begin{equation} 
Z_0=\sum_{\sigma_1,\sigma_2,\sigma_3=\pm 1} M_0^{\sigma_1\sigma_2\sigma_3} 
\end{equation}
where
\begin{equation} 
M_0^{\sigma_1\sigma_2\sigma_3}=\exp[-\beta E(\sigma_1,\sigma_2,\sigma_3)]
\end{equation}
$M_0^{\sigma_1\sigma_2\sigma_3}$ can take only 2 values because 
there are
only 2 classes of spin configurations with different energy:
\begin{equation} 
         \left\{ \begin{array}{l} 
         M_0^{\sigma\sigma\sigma}=e^{3 \beta J}=y^3  \\
         M_0^{\sigma\sigma(-\sigma)}=M_0^{\sigma(-\sigma)\sigma}
             M_0^{(-\sigma)\sigma\sigma}=e^{- \beta J}=y^{-1}
           \end{array}
   \right.
\end{equation}
where $y= e^{\beta J}$.
In terms of $y$ the partition function is:
\begin{equation} 
Z_0=2y^3+6y^{-1} 
\end{equation}

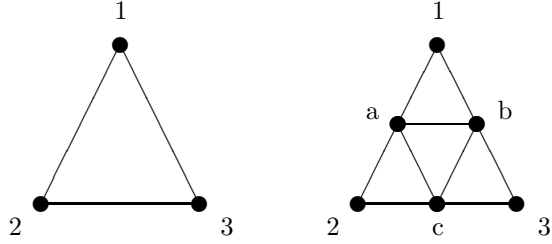
\begin{figure}[t]
  \begin{center}
  \begin{picture}(220,100)(-20,-20)
    \put(0,0){\begin{picture}(60,60) 
                      \multiput(0,0)(60,0){2}{\circle*{6}}
                      \put(30,60){\circle*{6}}
                      \put(0,0){\line(1,0){60}}
                      \put(0,0){\line(1,2){30}}
                      \put(30,60){\line(1,-2){30}}
                     \end{picture}}
    \put(28,70){1} \put(-12,-12){2} \put(68,-12){3}
    \put(120,0){\begin{picture}(60,60)
                   \multiput(0,0)(30,0){2}{\begin{picture}(30,30) 
                      \multiput(0,0)(30,0){2}{\circle*{6}}
                      \put(15,30){\circle*{6}}
                      \put(0,0){\line(1,0){30}}
                      \put(0,0){\line(1,2){15}}
                      \put(15,30){\line(1,-2){15}}
                     \end{picture}}
                   \put(15,30){\begin{picture}(30,30) 
                      \multiput(0,0)(30,0){2}{\circle*{6}}
                      \put(15,30){\circle*{6}}
                      \put(0,0){\line(1,0){30}}
                      \put(0,0){\line(1,2){15}}
                      \put(15,30){\line(1,-2){15}}
                     \end{picture}}
                  \end{picture}}
    \put(148,70){1} \put(108,-12){2} \put(188,-12){3}
    \put(123,32){a} \put(173,32){b} \put(148,-12){c}
  \end{picture}
  \end{center}
  \caption{Labeling of sites used for ${\cal G}_0$ and ${\cal G}_1$}
  \label{gasklab}
\end{figure}

For ${\cal G}_1$ the partition function can be expressed in the same way 
separating the sum over the states of the inner sites using the
tensor $M_1$ whose indices correspond to the spins on the external vertices
(fig. \ref{gasklab}):
\begin{equation} 
     M_1^{\sigma_1\sigma_2\sigma_3}=
     \sum_{\sigma_a,\sigma_b,\sigma_c=\pm 1} \exp[-\beta E(\sigma_i)] =
     \sum_{\sigma_a,\sigma_b,\sigma_c=\pm 1}
     M_0^{\sigma_1\sigma_a\sigma_b} M_0^{\sigma_a\sigma_2\sigma_c}
M_0^{\sigma_b\sigma_c\sigma_3}
\end{equation}
Now $M_1$ has the same structure as $M_0$ since the possible values are:
\begin{equation} 
           \left\{ \begin{array}{l} 
           M_1^{\sigma\sigma\sigma}=4y^{-3}+3y+y^9 \\
           M_1^{\sigma\sigma(-\sigma)}=M_1^{\sigma(-\sigma)\sigma}
             M_1^{(-\sigma)\sigma\sigma}=3y^{-3}+4y+y^5 
           \end{array} 
   \right.
\end{equation}
One can obtain $M_1$ from $M_0$ simply by a transformation mapping $y^3$
in $4y^{-3}+3y+y^9$ and $y^{-1}$  in  $3y^{-3}+4y+y^5$.
This is done by the substitution
\begin{equation}
y \to f(y)=\left( \frac{y^8-y^4+4}{y^4+3} \right)^{\frac{1}{4}}
\label{tr1}
\end{equation}
followed by the multiplication by
\begin{equation}
c(y)=\frac{y^4+1}{y^3} \left[ (y^4+3)^3(y^8-y^4+4)\right]^{\frac{1}{4}} 
\label{tr2}
\end{equation}
The transformation also gives the new partition function:
\begin{equation} Z_1(y)=Z_0(f(y))\ c(y) 
\end{equation}

Following the same argument one can obtain for the $(n+1)$th stage
of the gasket ${\cal G}_{n+1}$:
\begin{equation} 
Z_{n+1}(y)=Z_n(f(y))\cdot \left[c(y)\right]^{3^n} 
\end{equation}

Using this recursion relation we get:
\begin{equation} 
Z_n(y)=\frac{2}{y^{3^n}} P_n(y^4) 
\end{equation}
where $P_n(t)$ is a polynomial in $t$ of degree $3^n$ in which the
coefficient of $t^{3^n}$ is 1;
for $n=0$ one has $P_0(t)=t+3$
while the general case $n>0$ can be proven by induction.

\section{Zeros of the partition function}
Introducing the variable $x = y^4$ the transformation (\ref{tr1}), (\ref{tr2})
is given by:
\begin{equation} 
\left\{ \begin{array}{l} 
            x \rightarrow \tilde{f}(x)=\frac{x^2-x+4}{x+3} \\
               \tilde{c}(x)=(x+1)x^{-\frac{3}{4}}
                \left[(x+3)^3(x^2-x+4)\right]^{\frac{1}{4}} 
           \end{array} 
 \right.
\end{equation}
Denoting by $x_n^i$ the zeros of $P_n(x)$ the partition function reads
\begin{equation} 
Z_n=\frac{2}{x^{3^n/4}} \prod_{i=1}^{3^n}(x-x_n^i) 
\end{equation}
and using the recurrence one finds
\begin{equation} 
2 x^{-\frac{3^{n+1}}{4}} \prod_{i=1}^{3^{n+1}}(x-x_{n+1}^i)= 
   2 x^{-\frac{3^{n+1}}{4}} \prod_{i=1}^{3^n} \left\{ [(x^2-x+4)-x_n^i(x+3)]
        (x+1) \right\}
\end{equation}

This equation shows that for every root $x_n^i$ of $Z_n$, $Z_{n+1}$
has the root $x=-1$ and the 2 solutions of
\begin{equation} 
x_n^i=\tilde{f}(x_{n+1}^i) 
\end{equation}
namely the preimages of $x_n^i$ by the transformation $\tilde{f}$ .

Starting from $x_0^1=-3$
one obtains all the zeros of the partition function for the $n$-th stage
gasket as shown in table~\ref{zerij}, where $h(x)$ denotes the set of
the preimages of $x$ (and $h^k(x)$ is a set of $2^k$ zeros).
\begin{table}[ht]
 \[ \begin{array}{|c|l|}
      \hline
      n & \multicolumn{1}{c|}{\mbox{zeros}} \\ \hline \hline
      0& -3 \\  \hline
      1& -1\hspace{.5cm} h(-3) \\ \hline
      2& -1\hspace{.5cm} h(-1)\hspace{.5cm} h(h(-3)) \\ \hline
      3& -1\hspace{.5cm} h(-1)\hspace{.5cm} h(h(-1))\hspace{.5cm} 
            h(h(h(-3))) \\ \hline
      n& -1\hspace{.5cm} h(-1)\hspace{.5cm} \ldots \hspace{.5cm} 
            h^{n-1}(-1)\hspace{.5cm} h^n(-3) \\ \hline
   \end{array} \]
 \caption{Zeros of partition function in y}
 \label{zerij}
\end{table}
From this table one can see that the preimages of $-1$ by the $j$th iterate
of $\tilde{f}$ appear at the $(j-1)$th stage and are zeros of all the 
following stages, while the preimages of $-3$ are `temporary' zeros. 
It is also possible to find the multiplicity of these zeros: in fact since 
every
root generates the root $-1$ at the next stage, this value appears $3^{n-1}$
times among the zeros of $n$-th stage, the roots $h(-1)$ appear as many times 
as $-1$ in the previous stage (their multiplicity is $3^{n-2}$) and in general
the multiplicity of the roots belonging to $h^j(-1)$ is $3^{n-j-1}$.

In this way one can, in principle, calculate all the zeros of the 
partition function
at any stage. In practice this is possible only for small $n$ because of
their exponential growth.
An alternative approach \cite{derrida} is to start from a root 
$x_0$ (for example `$-1$' whose
preimages are `permanent') then choose at random one of its two preimages
by the transformation $\tilde{f}$ (denoted by $x_1$), then choose one of the
preimages of $x_1$ and so on; the set of points obtained in this way is a
representative of the set of all roots and has the advantage to contain zeros
relative to large $n$.

\begin{figure}
  \begin{center}
    \mbox{\epsfig{file=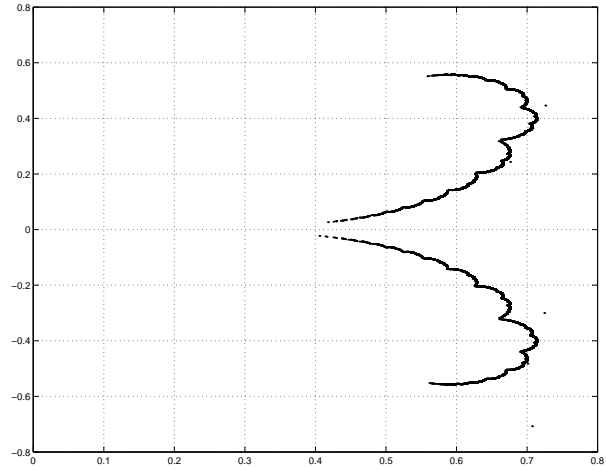,height=8cm,angle=90}}
  \end{center}
  \caption{20000 zeros obtained by the random method (in the plane of 
      $t=e^{-\beta J}$)}
  \label{rand}
\end{figure}

By plotting in the complex temperature plane the roots obtained with both
methods it can be seen that very few of them fall in the neighborhood of the
real axis and no information can be obtained about the critical behaviour (see
fig. \ref{rand} where the zeros are plotted in the plane of the variable 
$t=e^{-\beta J}$).

\begin{figure}
  \begin{center}
    \mbox{\epsfig{file=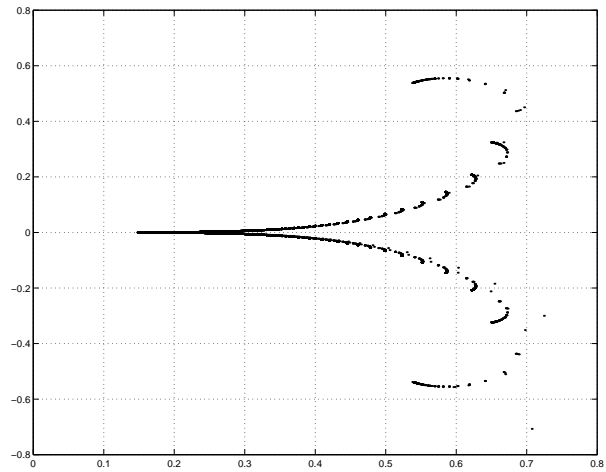,height=8cm,angle=90}}
  \end{center}
  \caption{20000 zeros obtained with probability 0.98 of choosing $h_2$ 
   (in the $t$ plane)}
  \label{rand98}
\end{figure}

A good technique to obtain more zeros near the real axis consists in changing the
choice probability  of the two preimages \cite{derrida}; the two solutions of
$x=\tilde{f}(x^{\prime})$ are
\begin{equation} 
x^{\prime}=h_1(x)=\frac{1 + x - {\sqrt{-15 + 14\,x + {x^2}}}}{2} 
\end{equation}
\begin{equation} 
x^{\prime}=h_2(x)=\frac{1 + x + {\sqrt{-15 + 14\,x + {x^2}}}}{2} 
\end{equation}
and one can see that the repeated application of $h_2$ gives a sequence
of points approaching the real axis. The set of roots obtained by increasing
the probability of choosing the second preimage is not a representative
set af all roots (it doesn't show their density not even approximately) but
gives us a chance to observe their behaviour in the interesting area (see
fig. \ref{rand98}).

\begin{figure}
  \begin{center}
    \mbox{\epsfig{file=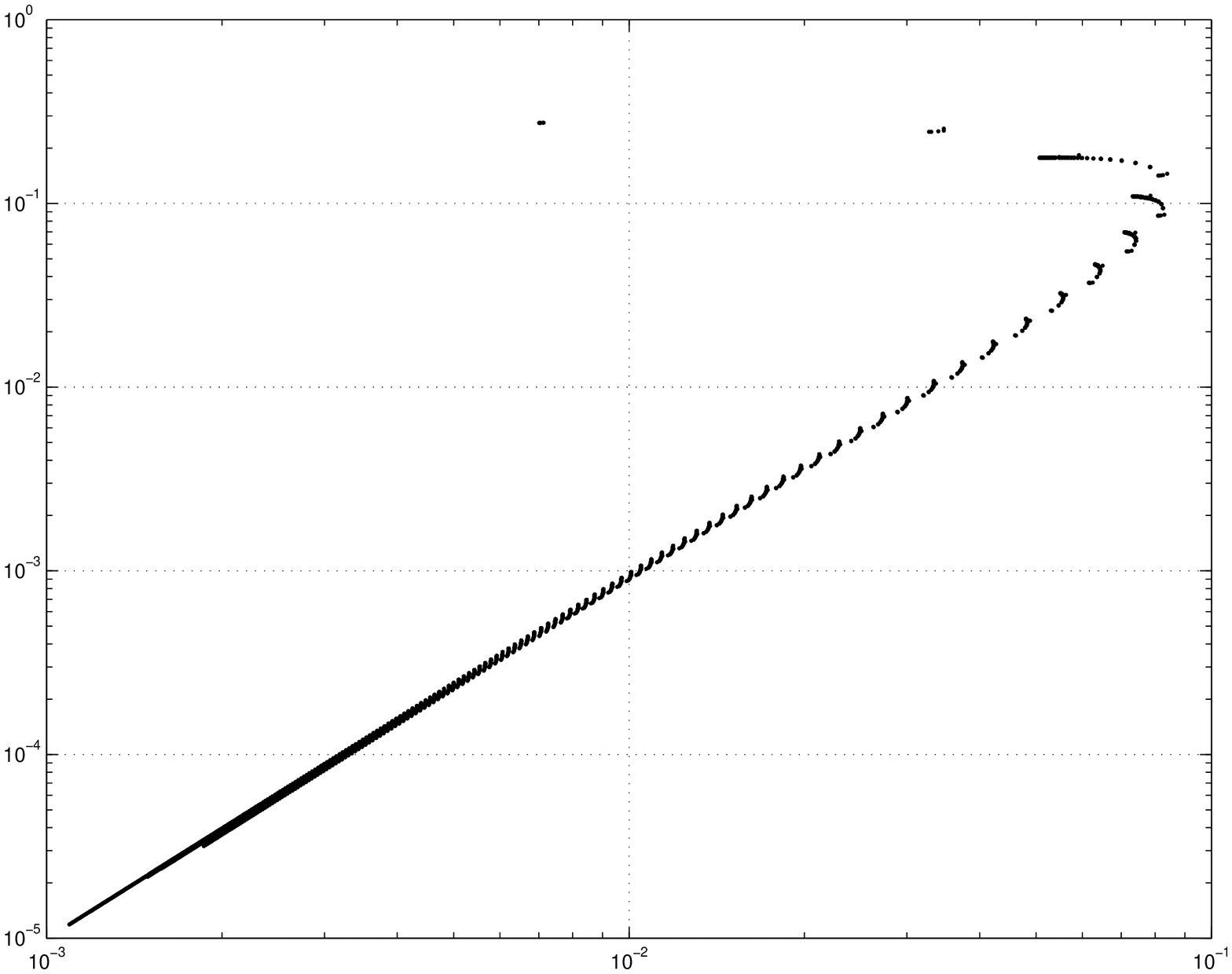,height=8cm,angle=90}}
  \end{center}
  \caption{Log-log plot of zeros (in the variable $w=e^{-4\beta J}$)}
  \label{loglog}
\end{figure}

A plot of these roots in the plane of $w=e^{-4\beta J}$ ($T=0$ corresponds to
$w=0$) with a {\it log-log} scale (fig. \ref{loglog}) shows that the real and
imaginary part are related by a power law: the curve can intersect the real
axis in the thermodynamic limit only at the origin.

Analytically one can verify this power behaviour by studying the
transformation of the variable $w=e^{-4\beta J}=x^{-1}$
\begin{equation} 
g(w)=\left. \frac{1}{f(x)} \right|_{x=\frac{1}{w}}=\frac{w(3w+1)}{4w^2-w+1}
\label{g}
\end{equation}
Assuming that $\Im(w)=A\ \Re(w)^b$, that is
\begin{equation} 
w=\xi+i A \xi^b 
\label{a}
\end{equation}
and inserting (\ref{a}) in (\ref{g}) one obtains
\begin{equation} 
\Im(g(\xi+i A \xi^b))= A \xi^b \left(1+8\xi+ O(\xi^{\min\{3,2b-1\}})\right)
\end{equation}
and
\begin{equation} 
A\ \left(\Re(g(\xi+i A \xi^b))\right)^b=A \xi^b \left(1+4b\xi
             O(\xi^{\min\{3,2b-1\}}))\right)
\end{equation}
Choosing $b=2$, one sees that the curve $w=\xi+i A \xi^2$ is `conserved'
by transformation $g$ except for higher order terms in $\xi$.

\section{Density of zeros}

We have seen that the zeros pinch the real axis only at $T=0$ and we
proceed by studying their density in the neighbourhood of this point 
to establish the critical behaviour; it is important to see whether
this density (which is quite small, as we have seen) goes to zero or
remains finite.
A numerical estimate can be obtained by simply counting the zeros (with
their multiplicity).

This has been done in two ways:
\begin{itemize}
 \item considering only the zeros in the neighbourhood of the real axis
   and grouping them with regard to their real part (one-dimensional density);
 \item dividing the complex plane in equal rectangles and counting the zeros
   contained (two-dimensional density).
\end{itemize}

\begin{figure}
  \begin{center}
    \mbox{\epsfig{file=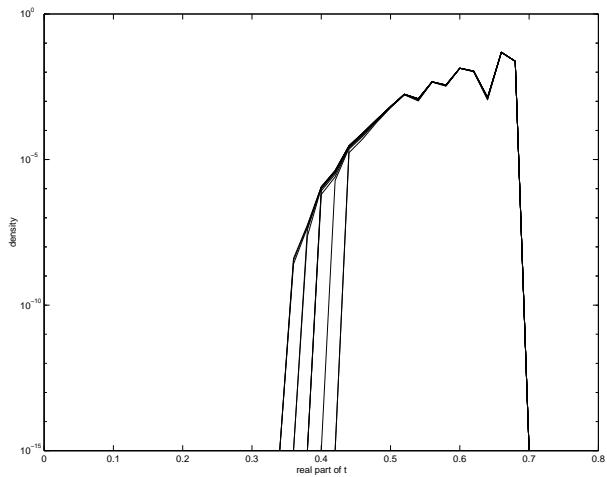,height=8cm,angle=90}}
  \end{center}
  \caption{Density of zeros vs. their real part in the $t$ plane (I method)}
  \label{dens1}
\end{figure}

\begin{figure}
  \begin{center}
    \mbox{\epsfig{file=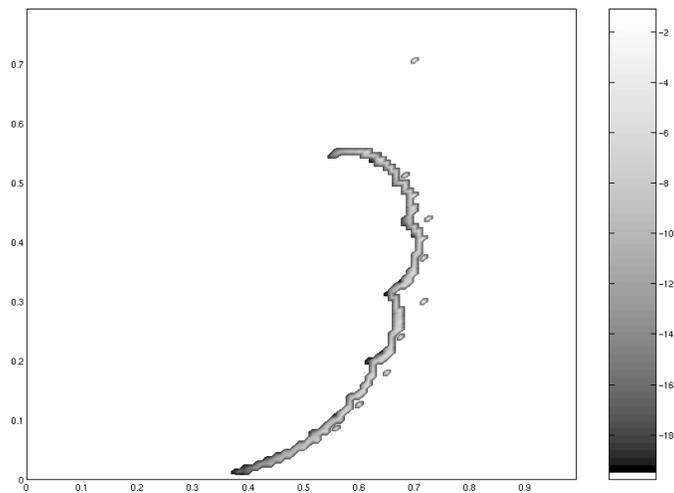,height=11cm,angle=270}}
  \end{center}
  \caption{Density of zeros in the $t$ plane (log scale)}
  \label{dens2}
\end{figure}

\begin{figure}
  \begin{center}
    \mbox{\epsfig{file=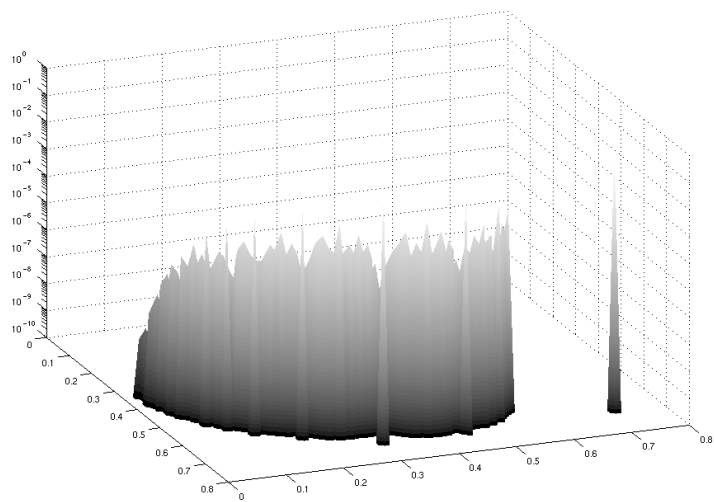,height=11cm,angle=270}}
  \end{center}
  \caption{3-dimensional view of the zeros density in the $t$ plane}
  \label{dens3}
\end{figure}

In the first case one obtains plots like fig. \ref{dens1} which
shows the density of the zeros of the partition function with 
$\Im(t)<.3$ for gaskets from ${\cal G}_{10}$ to ${\cal G}_{20}$; 
in the second case obtains the result shown in figures \ref{dens2} and 
\ref{dens3} (which refers to ${\cal G}_{18}$). 
From fig. \ref{dens1} one can
see that going from one stage to the next the density does not change 
appreciably except for the tail towards $0$ that grows longer but is
strongly decreasing: the density at $T=0$ appears to vanish exponentially.

This behaviour can also be verified by an analytical estimate. First we
notice that the zeros near $T=0$ are those obtained by the repeated
application of $h_2$.
Indeed for $|x|\rightarrow \infty$ we have
\begin{equation} 
h_1(x) \to k 
\end{equation}
while
\begin{equation} h_2(x) \to x+4-\frac{16}{x}+O(x^{-2}) 
\end{equation}
and, in terms of real and imaginary part,
\begin{equation} h_2(u+i v)\approx \left( u+4-\frac{16 u}{u^2+v^2}\right)+
        i \left(v-\frac{16 v}{u^2+v^2}\right)
\end{equation}

Applying $h_2$ to $z=u+i~v$ for large $u$ one obtains $h_2(z) \simeq 4u+i~v$.
In this limit the density of zeros in the $x$ plane becomes
the product of two factors, one depending only on $u$ and
the other on $v$:
\begin{equation} 
d(u,v) \approx d_1(u) d_2(v) 
\end{equation}
where $d_2(v)$ is bounded.

The asymptotic behaviour of $d_1(u)$ for $u\to\infty$ can be obtained
by noting that, for each set $h^k(-1)$, the
zeros with real part $u$ are those obtained by a sequence ending with 
the application of $h_1$ followed by $h_2^n$, where 
$n=\frac{u}{4}+c$ and $c$ is a constant independent of $u$. 
So this fraction of zeros is $\frac{1}{2}$ to the
power $\frac{u}{4}+c$, that is proportional to $\exp (-u/U)$ with
$U=4 \ln 2$.
Since the total density $d_1$ is a weighted sum of the partial densities that
have the same behaviour we have:
\begin{equation} 
d_1(u)\propto \exp(-u/U) 
\end{equation}

To find the density in the $t$ plane we must now divide $d$ by the Jacobian
of the transformation:
\begin{equation} 
t=x^{-\frac{1}{4}} 
\end{equation}
This Jacobian turns out to be
\begin{equation}
\frac{1}{16} (u^2+v^2)^{-\frac{5}{4}}
\end{equation}
and finally
\begin{equation} 
\tilde{d}(t_r,t_i)\propto\left. d_2(v) e^{-u/U}
        (u^2+v^2)^{\frac{5}{4}}  \right|_{t_r,t_i}
\end{equation}
where $t_r$ e $t_i$ are the real and imaginary part of $t$.

\begin{figure}[p]
  \begin{center}
    \mbox{\epsfig{file=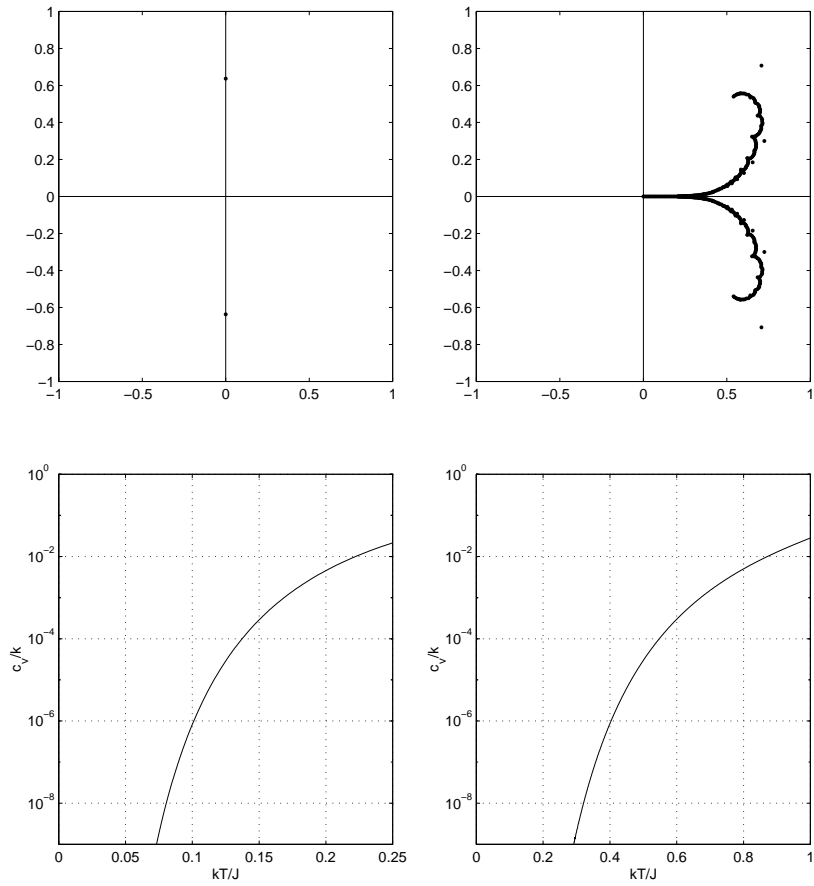,height=12cm}}
  \end{center}
  \caption{Comparison between the zeros distribution in the $t$ plane and 
the specific heat for the linear chain (on the left) and the gasket 
(on the right)}
  \label{comp}
\end{figure}

For $t\rightarrow0$ (that is $u\rightarrow +\infty$) the density
vanishes exponentially, as we could infer from numerical calculation,
and this behaviour has the same effect as a gap near the real axis.
Therefore the low temperature regime is not affected by the zeros contained in this
region and one observe a situation analogous to the one dimensional case.
This can be seen, for example, by comparing the behaviour of thermodynamical
quantities: 
figure \ref{comp} shows that, even if the zeros distributions 
seem to be quite 
different, the behaviour of the specific heat for the Sierpinski gasket and
the linear chain is essentially the same.

\section{Conclusions}
The anomalous behaviour of the density of zeros for the Ising model
on the Sierpinski gasket is to be deeply related to the its self-similar
geometry. This strongly suggests a careful approach to the analysis
of scaling of zeros density on fractals. In particular
a stimulating open problem is the relation of this scaling with the
geometry of a generic self-similar structure and with 
known anomalous dimensions. An important step in this direction would
be the study of Fisher zeros on the more complex case of a Sierpinski
carpet, where an exact solution is still lacking but the Ising
model is expected to have a phase transition at finite temperature.
 
\bibliography{}

\begin{thebibliography}{00}
\bibitem{panza1}Y. Gefen, B.B. Mandelbrot, A. Aharony, {\em Phys. Rev. Lett.} 
{\bf 45}, 855 (1980)
\bibitem{panza2}Y. Gefen, A. Aharony, Y. Shapir, B.B. Mandelbrot, {\em J.Phys. A}
{\bf 17}, 435 (1984)
\bibitem{mwg2}D. Cassi, {\em Phys. Rev. Lett.} {\bf 76}, 2941 (1996)
\bibitem{fss}R. Burioni, D. Cassi and A. Vezzani, Preprint UPRF-99-04, submitted
to J. Phys. A
\bibitem{sfer}D. Cassi and L. Fabbian, {\em J. Phys. A} {\bf 32},  L93 (1999)
\bibitem{on}R. Burioni, D. Cassi and C. Destri, cond-mat 9809334
\bibitem{leeyang1}C.N. Yang and T.D. Lee, {\em Phys. Rev.} {\bf 87}, 404 (1952)
\bibitem{leeyang2}T.D. Lee and C.N. Yang, {\em Phys. Rev.} {\bf 87}, 410 (1952)
\bibitem{ruelle}See i.e. D. Ruelle, {\em Statistical Mechanics}, W.A. Benjamin,
New York (1969)                    
\bibitem{ipz}C. Itzykson, R.B. Pearson, J.B. Zuber, {\em Nucl. Phys. B} {\bf 220}
[FS8], 415 (1983)
\bibitem{griff}R.B. Griffith, in {\em Phase transitions and critical phenomena}
Vol. 1, C. Domb and M.S. Green eds., Academic press, New York (1973)
\bibitem{fisher}M. Fisher, {\em Lectures in theoretical physics} Vol. VII C,
W.E. Brittin ed., Colorado Press, Boulder (1965)
\bibitem{shrock}V. Matveev and R. Shrock, {\em Phys. Rev. E} {\bf 53}, 254 (1996)
\bibitem{derrida}B. Derrida, L. De Seze and C. Itzykson, {\em J. Stat. Phys.}
                 {\bf 33}, 559 (1983)
\end{thebibliography}

\end{document}